\documentclass[twocolumn,amsmath,amssymb,prl,10pt,nofootinbib,superscriptaddress]{revtex4}

\usepackage{caption}
\usepackage[labelformat=simple]{subcaption}
\usepackage{graphicx}
\usepackage{dcolumn}
\usepackage{bm}
\usepackage{romannum}

\usepackage{amsmath}
\usepackage{amssymb}
\usepackage{float}
\usepackage[export]{adjustbox}
\usepackage[labelsep=period]{caption}
\usepackage[normalem]{ulem}
\usepackage{mathtools}
\usepackage{siunitx}
\usepackage{soul}
\usepackage{physics}
\usepackage{minitoc}

\usepackage{bm}
\usepackage{xcolor}
\usepackage{lipsum}

\begin{document}

\pagenumbering{arabic}

\preprint{APS/123-QED}

\title{ Search for black hole super-radiance using gravito-optic hetrodyne detection}

\author{Eduard Atonga}%
\email{eduard.atonga@univ.ox.ac.uk} 
\affiliation{%
Department of Physics, Atomic and Laser Physics sub-Department, Clarendon Laboratory, University of Oxford,\\
Parks Road, Oxford OX1 3PU, United Kingdom
}%

\author{Ramy Aboushelbaya}%
\email{ramy.aboushelbaya@physics.ox.ac.uk}
\affiliation{%
Department of Physics, Atomic and Laser Physics sub-Department, Clarendon Laboratory, University of Oxford,\\
Parks Road, Oxford OX1 3PU, United Kingdom
}%

\author{Peter A. Norreys}%
\email{peter.norreys@physics.ox.ac.uk}
\affiliation{%
Department of Physics, Atomic and Laser Physics sub-Department, Clarendon Laboratory, University of Oxford,\\
Parks Road, Oxford OX1 3PU, United Kingdom
}%
\affiliation{%
John Adams Institute for Accelerator Science, Denys Wilkinson Building, University of Oxford\\
Keble Road, Oxford OX1 3RH, United Kingdom
}%

\begin{abstract}

Gravitational-wave astronomy plays a crucial role in early universe cosmology, dark matter detection, and black-hole merger studies. It is shown here that the heterodyne detection of the gravito-optic effect (employing an enhanced Fabry-Pérot cavity) presents a novel broad band approach to gravitational wave detection, with favorable length and frequency scalings. It is demonstrated that such a device is well suited for detecting gravitational waves generated by the annihilation of bosons around black holes, enabling the exploration of a mass range from \( 2.40 \times 10^{-12} \, \mathrm{eV} \) to \( 4.54 \times 10^{-7} \, \mathrm{eV} \). Furthermore, due to the unique coherent nature of gravitational waves generated by boson annihilation any coherent gravitational waves detected above \( 1.23 \times 10^5 \, \mathrm{Hz} \) may be indicative of the existence of black holes of primordial origin.

\end{abstract}

\maketitle

The first successful detection of gravitational waves in 2016 by ground-based interferometric detectors LIGO, Virgo, and KAGRA, alongside the 2023 nanograv collaboration’s detection of the stochastic gravitational wave background using pulsar timing arrays, marked major milestones in gravitational wave astronomy \cite{PhysRevLett.116.061102,Lommen2017}. Upcoming observatories, such as LISA and Cosmic Explorer, aim to explore gravitational waves within the frequency regime between the aforementioned gravitational wave observatories \cite{Karsten_Danzmann_1996,Karsten_Danzmann_2003,PhysRevD.73.064030,PhysRevD.103.122004,hall2022cosmic,evans2021horizonstudycosmicexplorer}, leaving much of the spectrum beyond 10 kHz relatively unexplored. However, a range of potential gravitational wave signals have been predicted to exist above the LIGO frequency band covering several orders of magnitude, up to and above the optical frequency regime ($f\sim 10^{15} \text{ Hz}$), for which there are a variety of opportunities to discover new physics in the current universe and probe early universe cosmology in the time interval between the big bang and the emission of the cosmic microwave background due to the weak coupling of gravity to other fields and matter\cite{Aggarwal:2020olq}.

Over the decades, various detection schemes have been explored, including excitation of microwave resonator modes, conversion of gravitational waves into electromagnetic radiation in planetary magnetospheres \cite{PhysRevLett.132.131402}, and interactions with static magnetic and laser-based fields. A comprehensive review of these methods can be found in \cite{Aggarwal:2020olq}. Despite these extensive efforts, gravitational wave detection in these frequency bands remains elusive, as existing approaches either lack the necessary sensitivity or have yet to identify potential signals within the targeted range. 

A promising strategy for detecting gravitational waves beyond the 10 kHz frequency limit of LIGO and VIRGO involves levitated sensor detectors based on optically levitated multilayered dielectric microstructures. These interferometric detectors are designed to search for gravitational waves from primordial black holes and grand-unified-field-theory-scaled QCD axions produced through superradiance around black holes. However, their operational frequency range is constrained to 10–300 kHz \cite{PhysRevLett.110.071105,PhysRevLett.128.111101}, covering only a subset of the possible frequencies associated with these sources which is inversely proportional to the black hole mass and thus theoretically unbounded \cite{PhysRevD.81.123530,PhysRevD.83.044026}.

It is shown mathematically in \cite{Atonga_GravitoOptic}, that the gravito-optic effect, i.e. the frequency-shifted diffraction of light induced by a passing gravitational wave (analogous to acousto-optic diffraction \cite{Raman_Nath,https://doi.org/10.1002/j.1538-7305.1969.tb01198.x,Ghatak_Thyagarajan_1989}), serves as the basis for a novel detection scheme. This effect can be measured via optical heterodyning when the light–gravitational wave interaction is enhanced by a Fabry-Pérot cavity. In this Letter, a summary is provided for the operating principles of a detector based on the gravito-optic effect, along with its potential for detecting high-frequency gravitational waves produced by boson annihilation around black holes.

In the absence of an external current coupling to the electromagnetic field, the dynamics of electromagnetism in curved space-time with metric signature $(1,-1,-1,-1)$ is described by the minimally coupled Lagrangian \cite{Hobson, Maggiore:2007ulw}: 

\begin{equation}
    \mathcal{L}_{EM} = \frac{-1}{4\mu_0} g^{\mu \rho}g^{\nu \sigma}F_{\mu \nu}F_{\rho \sigma}  ,
\end{equation} 

where $g_{\mu \nu}$ is a symmetric tensor describing the curvature of space-time,  $F^{\mu \nu} = \nabla_{\mu} A_{\nu} - \nabla_{\nu}A_{\mu}$ is the anti-symmetric gauge invariant electromagnetic field tensor, which obeys the Bianchi identity $\nabla_{[\mu}F_{\rho \sigma]} = 0$, $A^{\mu}$ is denoted as the vector potential and  $\nabla_{\mu}$ is defined as the co-variant derivative. The associated equations of motion can be derived using the Euler-Lagrange equations. Assuming a torsionless space-time and using the Bianchi identity, after some algebra, the wave equation in curved space-time for the field tensor is \cite{bunney2021electromagnetism, park2021observation, tsagas2004electromagnetic}:

\begin{equation}
   \nabla_{\sigma} \nabla^{\sigma} F_{\mu \nu} + 2R_{\mu \sigma \nu \rho} F^{\sigma \rho} + R_{\mu}^{\sigma}F_{\nu \sigma} - R_{\nu}^{\rho}F_{\mu \rho} = 0, \label{BASE}
\end{equation}

where $R_{ab} = g^{cd}R_{acbd}$ and $R_{abcd}$ are the Ricci and Riemann tensors, respectively \cite{Hobson,Carroll}. 

In most circumstances, the solution for the full wave equation in curved space-time is typically intractable. However, for most applications, the length-scale variation of the metric $L_g \sim g/\partial g$ is much larger than the wavelength of the observed electromagnetic radiation $\lambda \ll L_g$, often denoted as the geometric optics limit. In this regime, the approximation leads to the well-known geodesic equation \( k^{\mu} \nabla_{\mu} k_{\nu} = 0 \) within the ray-optics formalism, which forms the fundamental basis of LIGO's operation. 

It is stressed here that the wave nature of light can still be retained in this approximation. Consider the case where space-times have been perturbed $g_{\mu \nu} = \eta_{\mu \nu} + h_{\mu}$ (with $|h_{\mu \nu}| \ll 1$) and where the perturbed metric satisfies the vacuum solutions of general relativity $R_{\mu \nu} = 0$. Then, by expanding the curved spacetime generalisation of the  D'Alembertian  operator ($\nabla_\sigma \nabla^\sigma$) in terms of partial derivatives to first order in the metric perturbation in the geometric optics limit (which is satisfied for the gravitational waves of interest from boson annihilation), the equation of motion of the electromagnetic field in the linear regime (ignoring terms $\mathcal{O}(h^2)$) can be written in terms of the electric field:

\begin{equation}
    (\Box - h^{\mu \nu}\partial_{\mu}\partial_{\nu})\bold{E}(t,\bold{r}) = 0 \label{DiffractionEOM}
\end{equation}

The dispersion relation in the geometric optics limit is $\mathcal{D}(\omega_0, \bold{k}_0) =  g_{\mu \nu}K_0^{\mu}K_0^{\nu} = 0$, where $K_0^{\mu}= (\omega_0/c, \bold{k}_0)$. The metric perturbation can be interpreted as an effective medium and is the basis of the so-called gravito-optic effect, in analogy to the acousto-optic effect \cite{Raman_Nath,https://doi.org/10.1002/j.1538-7305.1969.tb01198.x,Ghatak_Thyagarajan_1989}. 

During the interaction of an electromagnetic wave with a gravitational wave moving along the z-axis (where the metric in the laboratory frame is written in the Fermi-normal coordinate system \cite{MinserFNC,Fort}):

\begin{gather}
    \bar{g}_{00} = 1-  R_{0k0l} x^kx^lz^n \\
    \bar{g}_{0j} = -\frac{2}{3} R_{0kjl} x^kx^lz^n \\
    \bar{g}_{ij} = \eta_{ij}-\frac{1}{3} R_{ikjl}x^kx^lz^n,
\end{gather}

and the gauge-invariant Riemann tensor is computed from the gravitational wave described in the transverse traceless gauge $    h_{\mu \nu}^{TT} = h_{+}e^{+}_{\mu \nu}cos(\Omega t - K z) $.  

Successive diffraction orders $\omega_q = \omega_0 + q\Omega$ where $q \in \mathbb{Z}$ are generated and the plane waves satisfy the dispersion relation $\omega_{q} = c k_q$. However, because the higher diffraction orders $|q| \geq 2$ are of $\mathcal{O}(h^2)$ or greater, in the linear regime, one is limited to the lowest-order diffraction orders $q = \pm 1$.  

Now examine the limiting case in which the electromagnetic wave propagates orthogonally to an incident gravitational wave. In this case, the electromagnetic waves move along the positive and negative x-axis. Now consider the simplest case and choose the incident electric field vector $\bold{E}_0$ to be orthogonal to the wave vector of the incident gravitational wave. It follows that the solution to equation (\ref{DiffractionEOM}) with incident backward and forward propagating waves is written as the sum of scalar waves:

\begin{eqnarray}
  E_f(t,\bold{r}) = \sum_{q\in \{0,\pm 1\}} \mathcal{E}_f^{(q)}(\bold{r})e^{i(\omega_q t - \bold{\chi}_q\cdot \bold{r} )} \\ 
  E_b(t,\bold{r}) = \sum_{q\in \{0,\pm 1\}} \mathcal{E}_b^{(q)}(\bold{r})e^{i(\omega_q t + \bold{\chi}_q\cdot \bold{r} )},
\end{eqnarray} 
where \( \mathcal{E}_f^{(q)} \) and \( \mathcal{E}_b^{(q)} \) denote the electric field amplitudes of diffraction order \( q \) generated by the forward and backward-propagating background waves, respectively, and \( \mathcal{E}_f^{(0)} = \mathcal{E}_q^{(0)} = E_0 \).  The propagation vector of each order \( q \) is given by \( \boldsymbol{\chi}_q = (\boldsymbol{\alpha}_q, 0, -qK) \), where  \( \boldsymbol{\alpha}_q \) corresponds to the wave number of diffraction order \( q \) along the propagation direction of the background plane waves.   Defining the Helmholtz operator $\Delta^{(\pm)} = k_{\pm}-\nabla^2$, for a laser beam whose interaction length $L$ is much larger than its width $w_0$, the resulting Helmholtz equations for the lowest diffraction orders are given by:

\begin{eqnarray}
   \Delta^{(\pm)}\bold{\mathcal{E}}_f^{(\pm)}(\bold{r}) e^{-i\alpha_{\pm}x} = \frac{E_0 h_+ k_0^2 K^2}{4}x^2 e^{-i(\bold{k}_0 \pm \bold{K}) \cdot \bold{r}}  \\
     \Delta^{(\pm)}\bold{\mathcal{E}}_b^{(\pm)}(\bold{r}) e^{i\alpha_{\pm}x} = \frac{E_0 h_+ k_0^2 K^2}{4}x^2 e^{i(\bold{k}_0 \mp \bold{K}) \cdot \bold{r}}.
\end{eqnarray}

Now, define the function $\Lambda_\pm(L)$: 

\begin{eqnarray}
     \Lambda_\pm(L) =    \left( \frac{L^2}{\Delta \alpha_\pm} - \frac{2}{\Delta \alpha^3_\pm}\right) \text{sin}(\Delta \alpha_\pm L) \nonumber \\ 
    + \frac{2L}{\Delta \alpha^2_\pm} \text{cos}(\Delta \alpha_\pm L),
\end{eqnarray}

where $\Delta \alpha_{\pm} = k_0 -\alpha_{\pm} $. It follows that solutions for the electric field amplitude of each diffraction are given by: 

\begin{eqnarray}
    E^{\pm}_{f}(t,\bold{r}) \approx \frac{i E_0 h_+ k^2_0 K^2}{16 \alpha_{\pm}} e^{i(\omega_{\pm} t -  \alpha_{\pm} x \mp K z)}\Lambda_\pm(L)  \\ 
     E^{\pm}_{b}(t,\bold{r}) \approx \frac{-i E_0 h_+ k^2_0 K^2}{16 \alpha_{\pm}} e^{i(\omega_{\pm} t + \alpha_{\pm} x \mp K z)}\bar{\Lambda}_\pm(L),
\end{eqnarray}

where $\alpha_{\pm} = \sqrt{k^2_{\pm} - K^2}$. It immediately follows that the background forward and backward propagating electromagnetic plane waves interacting with a gravitational wave generate two diffraction orders with frequencies $\omega \pm \Omega$ at diffraction angles $\theta_\pm$ (albeit in opposite directions along the x-axis, respectively). 

The diffraction angle itself is computed from the relation: 

\begin{equation}
    \text{cos}(\theta_{\pm}) = \frac{\alpha_{\pm}}{k_{\pm}}.
\end{equation}

In the geometric optics limit, the diffraction angle can be approximated by $\theta_\pm \approx \Omega / \omega_0 \ll 1$.

The amplitude of the generated diffraction pattern and the diffraction angle are too small to produce a measurable and distinguishable signal unless extremely large propagation distances from the interaction region are considered. 

\begin{figure}[h]
\centering
\includegraphics[scale=0.28]{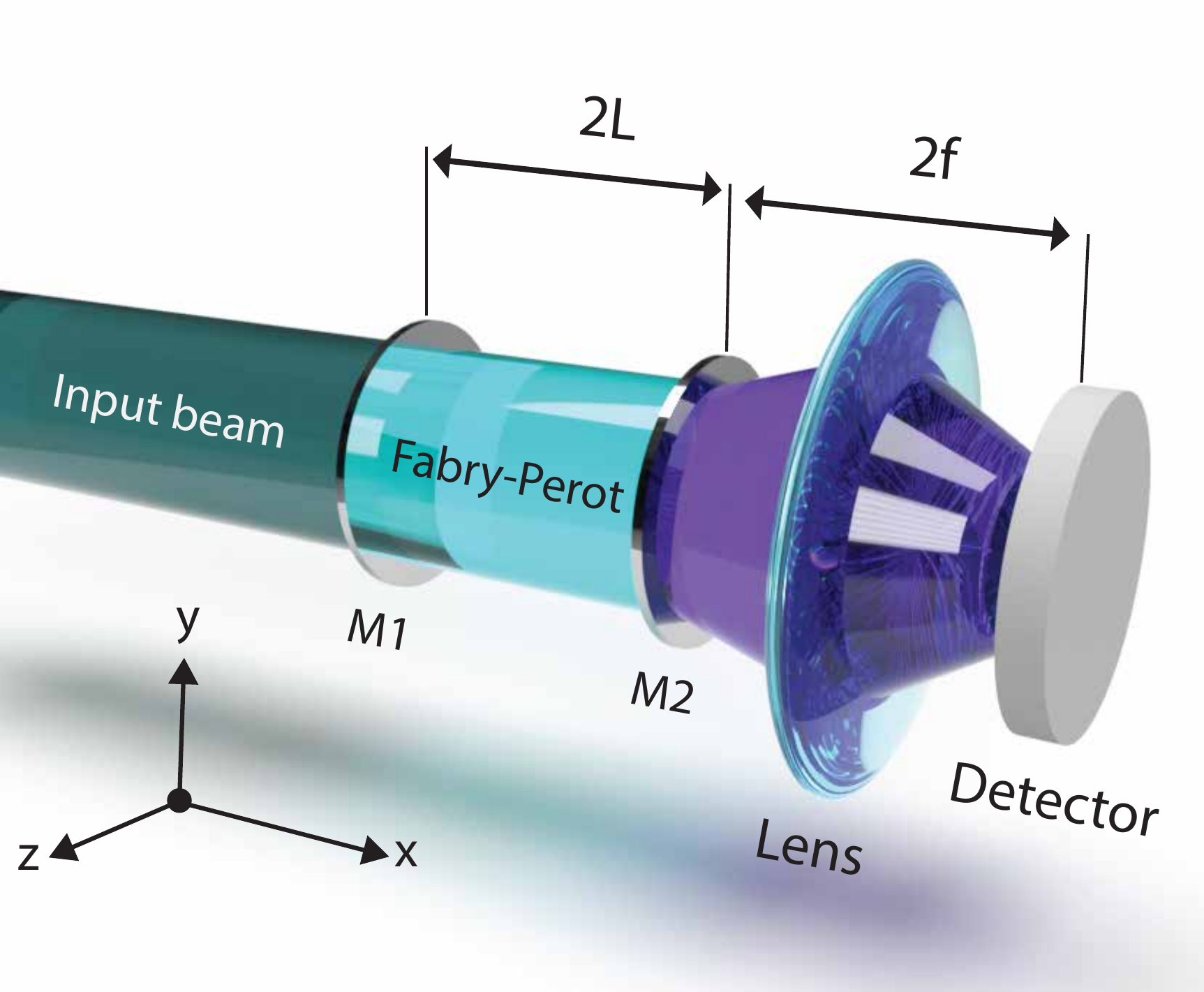}
\caption{Depiction of detection scheme geometry, describing the interaction of a incident electromagnetic beam propagating in the positive x-direction with a gravitation wave propagating in the positive z-direction. The signal  is imaged by a \textit{2f} lens geometry with $f \ll L$.}
\label{fig:tube}
\end{figure}

However, it is possible to obtain a measurable signal with a reasonable optical arrangement by: first, enhancing the effect through a Fabry-Pérot cavity; second, taking advantage of the shallow diffraction angle $\theta_{\pm}$ to measure the frequency shift of the diffraction orders with respect to the background electromagnetic beam. via heterodyne measurement of the beat frequency. An experimental setup illustrating this concept is shown in Figure (\ref{fig:tube}).

With each reflection of the background beam, the newly generated diffraction order experiences a different initial phase of the gravitational wave. Taking this into account when computing the interference of all electromagnetic signals inside the Fabry-Pérot cavity, in the limit where the observation time $T$ is much greater than the cavity build-up time the signal-to-noise ratio for the heterodyne signal \cite{PhysRevD.104.L111701,HALLAL2022100914,10.1119/1.15655,narayanan2002high} measured by the detector can be written in terms of the spectral noise density 
$S/N = \sqrt{ h^2_+ T/ S_\Omega }$. The expression for the spectral noise density is given by:

\begin{eqnarray}
    S^{\frac{1}{2}}_\Omega = \frac{(1-R)}{k^2_0K^2}\sqrt{\frac{\hbar}{2 P_0} \sum_{j\in \pm} \frac{\omega_j}{\Lambda^2_j(L) \mathcal{G}_j(w_0, L) }},
\end{eqnarray}

where the integral function is defined as:   

\begin{gather}
      \mathcal{G}_\pm(w_0, L) = \frac{1}{\pi} \int_{\mathbb{R}} \frac{\text{sinc}^2(\tau)\text{sin}^2(\gamma_\pm(\tau)/2)}{1+\mathcal{F} \text{sin}^2(\psi_\pm(\tau))} d\tau, \\ 
      \gamma_\pm(\tau) = 2 k_\pm L \pm 2 \tau \frac{L}{w_0}\text{sin}(\theta_{\pm}) \\ 
     \psi_\pm(\tau) = 4 k_\pm L(\text{cos}(\theta_\pm)-1) \pm 4 \tau \frac{L}{w_0}\text{sin}(\theta_{\pm}),
\end{gather}
 
with $ \mathcal{F} = 4R/(1-R)^2$ and $R$ is the reflectivity of the Fabry-Pérot cavity mirrors.

Having established the detection mechanism, the reader's attention is now drawn to the physics in which bosonic fields are confined to regions around a black hole, forming bound Keplerian levels \cite{Ternov1978,Damour1976,ZOUROS1979139,PhysRevD.22.2323}. This configuration, often referred to as a "gravitational atom," exhibits parallels to atomic systems such as a hydrogen atom, including the emergence of an analogous "fine-structure constant" $\alpha$ \cite{PhysRevD.81.123530,PhysRevD.83.044026, PhysRevD.91.084011, PhysRevD.95.043001, PhysRevD.104.103009}:

\begin{equation}
    \alpha = \frac{G M_{BH} \mu_a}{\hbar c^3},
\end{equation}

where $M_{BH}$ and $\mu_a$ are the black hole mass and the rest-mass energy of the axion, respectively. These bound states, similar to those of the hydrogen atom, are characterized by associated principal, orbital, and magnetic numbers $\{n,l,m\}$, as well as discrete energy levels:

\begin{equation}
E_n  = \mu_a \Bigl( 1- \frac{\alpha^2}{2n^2} \Bigr).
\end{equation}

It is interesting to note that massive bosonic waves are amplified around rotating black holes through the Penrose superradiance process \cite{PhysRevD.91.084011, PhysRevD.95.043001,PhysRevD.96.035019,PhysRevD.103.095019}. The occupation numbers of the bound levels grow exponentially by extracting energy and angular momentum from the black hole \cite{Aggarwal:2020olq, PhysRevD.91.084011, PhysRevD.95.043001}, when the super-radiance condition is satisfied \footnote{it is assumed throughout this letter that $l =m$.}: 

\begin{equation}
    \alpha \leq \frac{l}{2}\frac{a_*}{1+\sqrt{1-a_*^2}},
\end{equation}

where $a^*$ is the dimensionless spin of the black hole and the growth of the occupation number of the energy levels stops when enough angular momentum has been extracted so that the super-radiance condition is no longer satisfied, which occurs when a sufficient difference in the black hole spin has been extracted. 

Bosons within the bound states surrounding the black hole generate gravitational waves in a process analogous to electron-positron annihilation into photons in the presence of a nuclear background, through annihilation into gravitons in the black hole background ($a + a \rightarrow h $). The superradiance condition imposes an upper bound on the parameter ratio $\alpha/l\leq 1/2$, corresponding to $a^*=1$. Considering this constraint and adopting the relationship $n=l+1$, one can demonstrate that the bound-state energy level approaches a limiting value in the large-$l$ limit $\lim_{l \to \infty} E_l = 0.875\mu_a$. For smaller values of $l$, however, each axion bound-state energy represents a larger fraction of the axion rest mass, suggesting that the gravitational wave frequency $\Omega$ associated with the emitted radiation can be approximated as being independent of the fine-structure constant $\hbar \Omega \approx 2\mu_a$, reducing the parameter space for a preliminary investigation. The corresponding amplitude of the generated gravitational wave has been estimated \cite{
Arvanitaki:2014wva}:

\begin{equation}
    h\sim 10^{-22}\Bigl(\frac{1\text{ kpc}}{r} \Bigr)\Bigl(\frac{\alpha /l}{0.5} \Bigr)^\frac{p}{2} \frac{\alpha^{-\frac{1}{2}}}{l}\Bigl(\frac{M}{10 M_{\odot}} \Bigr)
\end{equation}

where $M_\odot$ denotes the solar mass, $r$ is the distance from the detector, $p= 17$ for $l = 1$ and $p= 4l +11$ for $l \geq 2$.

Now consider gravitational wave signals emitted by black holes within the Milky Way galaxy, specifically considering scenarios where the detector is precisely aligned with the source. In this context, one focuses on three black holes: Cygnus X-1; LMC X-1; and GRO J1655-40. All are located within 50 kpc of Earth, have masses ranging from 5 to 21 solar masses, and have spin measurements known to a certainty $3\sigma$ \cite{10.1046/j.1365-8711.2002.05189.x,Hyde_2017,10.1093/mnras/staa2656,ZHU2024381,Orosz_2011,Gou_2014,Zhao_2021}, as detailed in Table \ref{Tab:BHparameters}.

\begin{table}[h]
\begin{tabular}{|l|l|l|l|}
\hline
Name         & Mass ($M_{\odot}$) & Distance (kpc) & $a^*$ \\ \hline
Cygnus X-1   & 21                                                 & $2.25 \pm 0.08$      & \textgreater{}0.983  \\ \hline
LMC X-1      & $10.91 \pm 1.41$                                       & $ 48.10 \pm 2.22$   & \textgreater{}0.95   \\ \hline
GRO J1655-40 & $5.31 \pm 0.07$                                          & $3.7\pm 1.1$        & \textgreater{}0.983  \\ \hline
\end{tabular}
\caption{Black hole parameters.}
\label{Tab:BHparameters}
\end{table}

Since the gravitational wave signal is long-lived and monochromatic, the natural measure of signal strength is the amplitude of the sinusoidal wave. The signal-to-noise ratio builds up over extended integration times, allowing the strain sensitivity to be expressed in terms of the spectral noise density as $\sqrt{S^{\frac{1}{2}}_\Omega /T}$ \footnote{This expression is specific to monochromatic signals and does not directly apply to other types of signals, such as inspiraling or stochastic sources. However, the spectral noise density remains a useful quantity for calculating the sensitivity to these signals \cite{Aggarwal:2020olq, Moore_2015}.}. 

\begin{figure}[H]
\centering
\hspace{-0.5cm}
\includegraphics[scale=0.7]{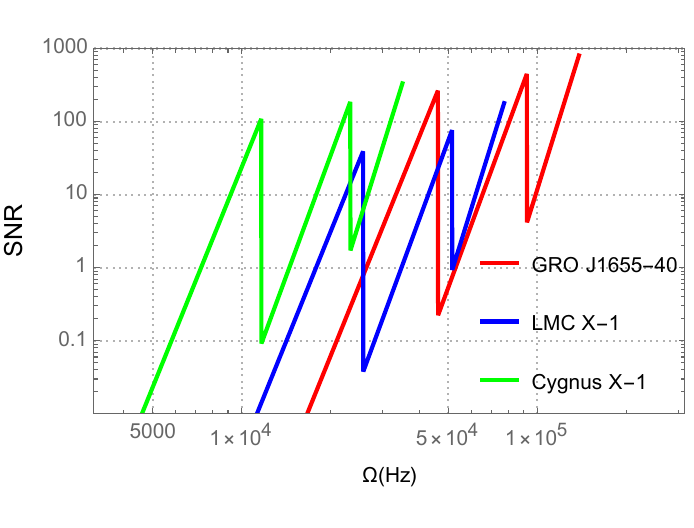}
\caption{combined SNR curves for $l = 1,2,3$ for a 2 km Fabry-Perot cavity of finesse $\mathcal{F} = 10^4$ for a observation time of $10^6$s.}
\label{fig:SNR}
\end{figure}

To evaluate the detectability of the beat frequency induced in the detector by the diffracted sidebands (generated through the space-time periodic modulation of the effective refractive index caused by the gravitational wave), the relevant experimental parameters are outlined here. The input laser power is set at $P_0 = 1 \text{ W}$,  two orders of magnitude below the input power of Advanced LIGO, which operates at $125 \text{ W}$ and is frequency-stabilized to deviations below 1 Hz over timescales pertinent to gravitational wave detection \cite{fritschel2001advanced, Izumi:14}. The distance between the collection lens and the detector is $f= 1 \text{ m} $. Conventional laser mirrors typically have reflection coefficients of 0.99, corresponding to a cavity finesse of a few hundred. However, optical supermirrors, typically dielectric in nature, can be optimized for exceptionally high reflectance. Fabry-Pérot cavities constructed with such mirrors achieve finesse values as high as $\mathcal{F} = \mathcal{O}(10^6)$ in the mid-infrared to optical frequency spectrum, enabling highly sensitive detection capabilities \cite{sones1997optimization,Truong,Muller:10,Meng:05}. Thus, for different values of the cavity length $L \gg f$, the signal-to-noise ratio (SNR) of the combined $l=1,2,3$ signal for each black hole is given in figure (\ref{fig:SNR}),

From figure \ref{fig:SNR} it can be seen that the proposed \(2 \, \text{km}\) detector operating over a \(10^6 \, \text{s}\) observation period (just under 12 days of continuous operation) enables exploration of axion mass ranges spanning three orders of magnitude, even when considering only the selected black holes. Higher orbital angular momentum levels allow for probing larger axion masses for a given black hole, while the sharp drop in the signal occurs when the preceding \(l\)-value fails to satisfy the superradiance condition.

The only astrophysical reliable source of black hole formation is via the collapse of stars. Whether or not a star collapses into a black hole depends on whether the star is more massive than the Tolman–Oppenheimer–Volkoff limit. Earlier estimates of this limit ranged between \(2.2\) to \(2.9 \, M_\odot\) \cite{bombaci1996maximum,Kalogera_1996}. However, further refinement due to the first gravitational wave detection from merging neutron stars GW170817 and, accounting for the rotation of the progenitor star, the lower limit on the mass of the resulting black hole is now estimated to be \(2.6 \, M_\odot\) \cite{Pooley_2018, doi:10.1126/science.359.6377.724, Rezzolla_2018}. Gravitational waves emitted by the annihilation of bosons in bound states with orbital angular momentum and dimensionless spin \( l=1 \) and \( a^*=1 \), respectively at a black hole mass of \(2.6 \, M_\odot\) would have a frequency of $\Omega_L = 1.22 \times 10^5 \text{ Hz}$. Under the condition $\hbar \Omega \approx 2\mu_a$, the black hole mass and orbital angular momentum decouple from the frequency of the emitted gravitational waves. However, by the supperradaince condition there exists a upper bound on the axion mass below which occupation numbers for the bound-state levels grow. As a consequence, there exists a maximal limit on the gravitational wave frequency that can be produced through axion annihilation processes:

\begin{equation}
    \Omega \leq\frac{l c^3}{GM_{BH}}
\end{equation}

\begin{figure*}[t!]
    \centering
    \begin{subfigure}[b]{0.3\textwidth}
        \centering
        \includegraphics[width=\linewidth]{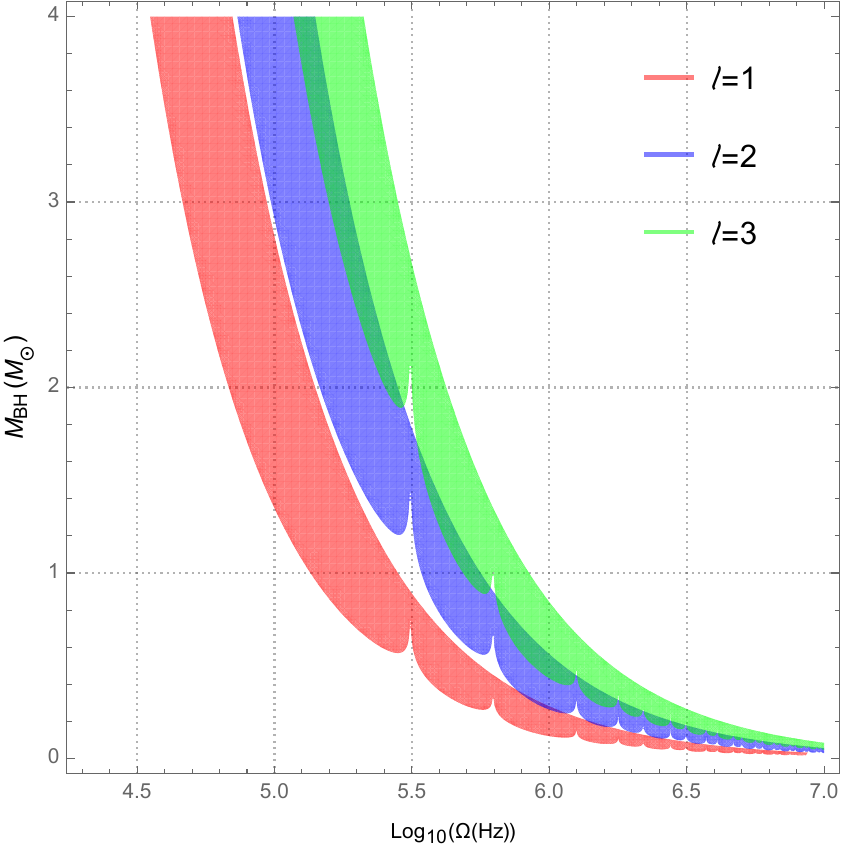}
        \caption{$10^4 \text{ Hz}\leq\Omega\leq 10^7 \text{ Hz}$}
    \end{subfigure}
    \hfill
    \begin{subfigure}[b]{0.3\textwidth}
        \centering
        \includegraphics[width=\linewidth]{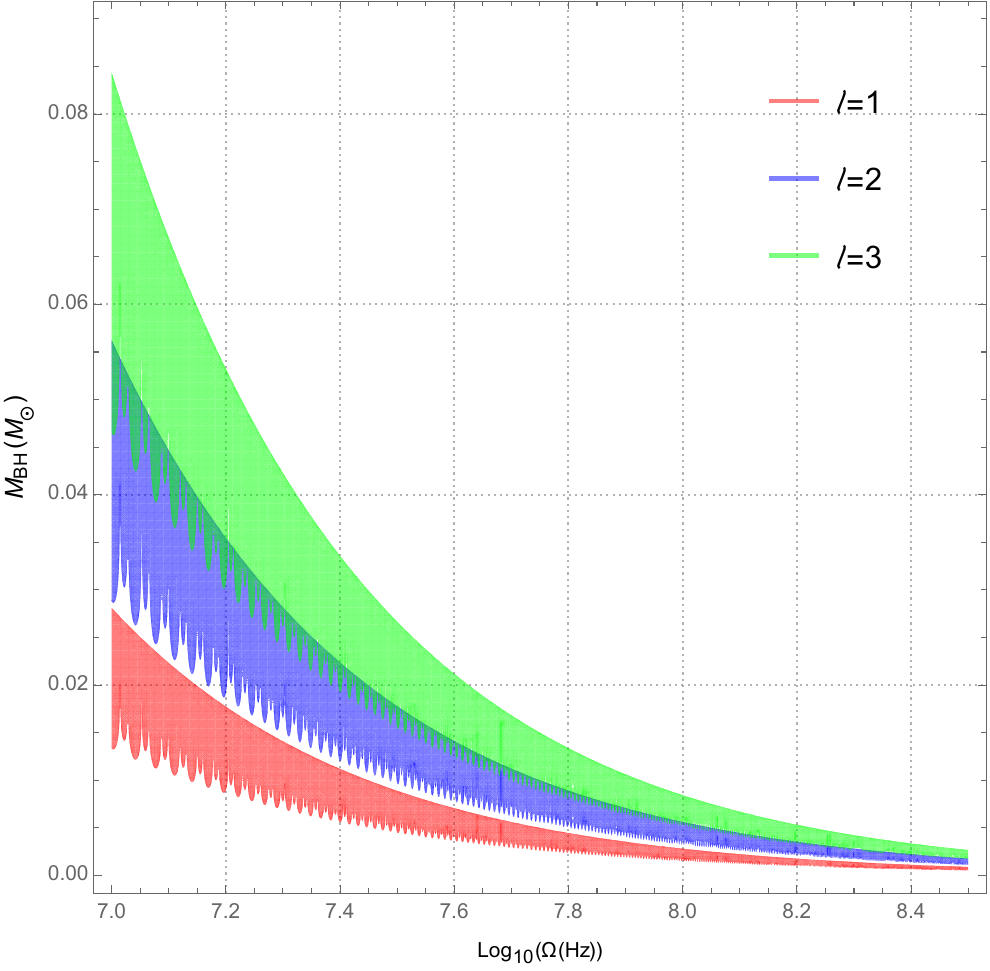}
        \caption{$10^7 \text{ Hz}\leq\Omega\leq 10^{8.4} \text{ Hz}$}
    \end{subfigure}
    \hfill
    \begin{subfigure}[b]{0.3\textwidth}
        \centering
        \includegraphics[width=\linewidth]{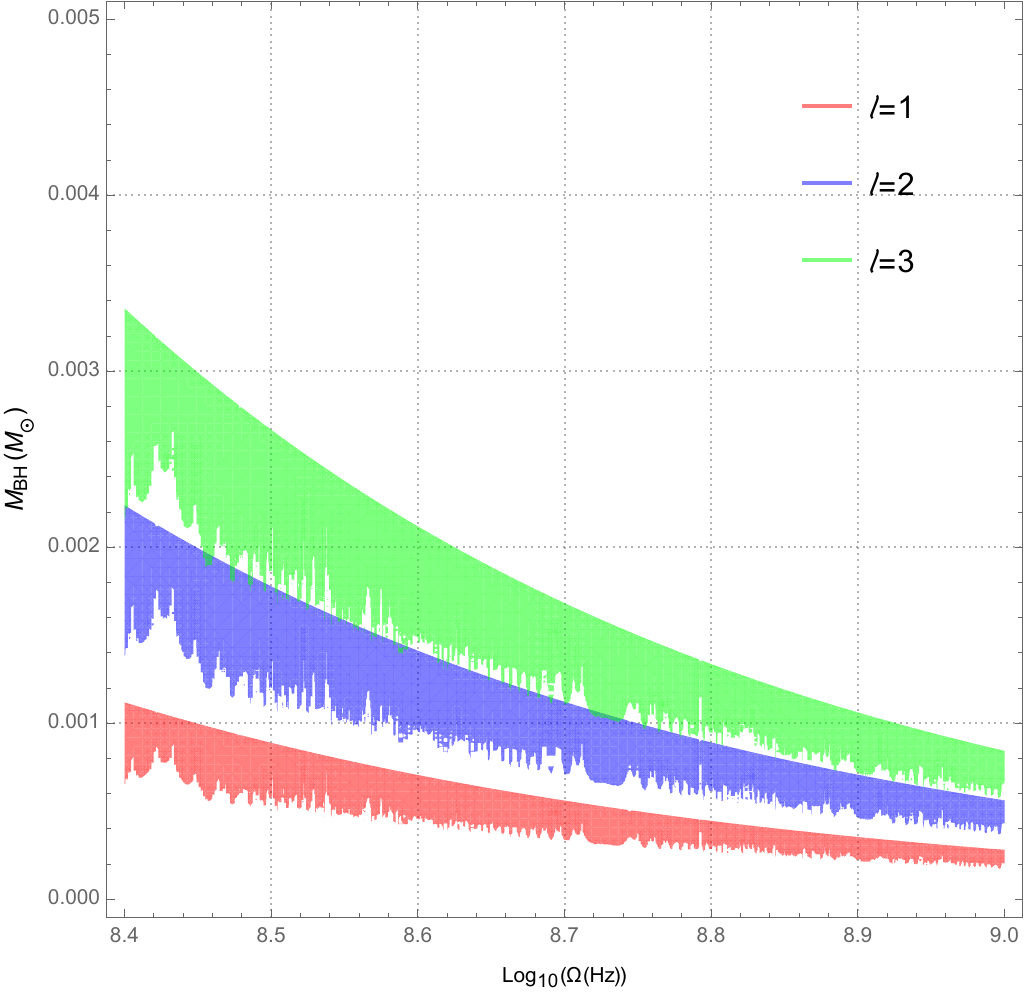}
        \caption{$10^{8.4} \text{ Hz}\leq\Omega\leq 10^9 \text{ Hz}$}
    \end{subfigure}
    \caption{SNR region plot for a 2Km fabry perot cavity of finesse $\mathcal{F} = 10^4$ for a observation time of  $10^6$s for different orbital angular momentum values for black holes of varying mass at a distance $r = 50 \text{ kpc}$.}
   \label{fig:PBH1}
\end{figure*}

Thus lower mass black holes from bosons in bound states with higher orbital angular momentum, emit higher frequency gravitational waves. The detection of any frequency exceeding $\Omega_L$ would then hint at the existence of black holes of masses below \(2.6 \, M_\odot\), which must be of primordial origin \cite{Maxim,PhysRevD.94.083504,Carr_2021}, if the orbital angular momentum could be determined. Figure \ref{fig:PBH1} illustrates the regions where the signal-to-noise ratio (SNR) exceeds 1, across a range of black hole and axion masses.
\par\vspace{1em}
\onecolumngrid
\twocolumngrid

In summary, in this letter a novel broadband search strategy has been presented. It is one that demonstrates the potential of a gravito-optic heterodyne detector as a viable method for probing bosonic fields that potentially exist in the vicinity of astrophysical or primordial black holes. In addition, higher boson masses can be investigated by targeting lower-mass black holes or considering higher orbital angular momentum numbers \( l \) for a given black hole. As explicit examples, the black holes LMC X-1, Cygnus X-1, and GRO J1655-40 have been analyzed, where the accessible boson mass ranges from \( 2.40 \times 10^{-12} \, \mathrm{eV} \) to \( 4.54 \times 10^{-11} \, \mathrm{eV} \).

The primary complementary gravitational wave detector for the observation of axion superradiance is the elegantly proposed Leviation-Sensor-Device (LSD), which operates in the frequency range \( f_{\mathrm{GW}} = \Omega/2\pi \) spanning from 10 kHz to 300 kHz. The upper frequency bound corresponds to a black hole mass of \( 0.12 \, M_\odot \), lying below the Tolman–Oppenheimer–Volkoff limit of \( 2.6 \, M_\odot \), thus enabling the exploration of the existence of primordial black holes. It is highlighted here that the new gravito-optic heterodyne detector is capable of operating over a much broader frequency range, constrained only by the electronic response rate of 10 GHz. At this frequency, the signals correspond to a primordial black hole mass of approximately \( 2.03 \times 10^{-4} \, M_\odot \) (for \( l=1 \) and \( a^*=1 \)) and a boson mass of \( 3.29 \times 10^{-7} \, \mathrm{eV} \).

It is emphasized here that the construction and operation of a gravito-optic heterodyne detector requires a highly stable laser source and precisely controlled cavity lengths, to ensure that any signal measured are solely attributable to the gravito-optic effect. Fortunately, these requirements are well within the technological capabilities demonstrated by current and next-generation interferometric gravitational wave detectors. Advanced LIGO, for example, achieves a fractional arm-length change due to noise smaller than \(\Delta L / L < 10^{-23}\) and maintains fractional frequency deviations below 1 Hz. Cosmic Explorer is designed to meet or surpass these benchmarks, featuring arm lengths exceeding \(40 \, \text{km}\) and circulating power of \(125 \, \text{MW}\) within its Fabry-Pérot cavities \cite{hall2022cosmic,PhysRevLett.134.051401}. Additionally, such precision is maintained over entire observing runs, which can persist for several months to a year — far exceeding the estimated observation time required for a detection in the scheme proposed here. While the LSD device imposes similar requirements (but additionally necessitates the levitation of a dielectric disk), the gravito-optic heterodyne detector offers a simpler alternative for detecting gravitational waves at frequencies exceeding the 10 kHz limit of LIGO.

\section{Acknowledgements} 

The authors gratefully acknowledge useful discussions with Dr. Aur\'elien Barrau and Dr. Killian Martineau from Université Grenoble Alpes and members of the Norreys' research group in the Clarendon Laboratory. They also thank all of the staff of the Central Laser Facility, Rutherford Appleton Laboratory for their assistance in the development of this work. This research was funded in whole or in part by the Oxford-ShanghaiTech collaboration agreement and UKRI-STFC grant ST/V001655/1. For the purpose of Open Access, the authors have applied a CC BY public copyright licence to any Author Accepted Manuscript (AAM) version arising from this submission.

\bibliography{refs.bib}

\end{document}